\documentclass[conference,compsoc]{IEEEtran}
\ifCLASSOPTIONcompsoc
  \usepackage[nocompress]{cite}
\else
  \usepackage{cite}
\fi

\ifCLASSINFOpdf
\else
\fi
\usepackage{bm,graphicx,mathrsfs,amsmath,amssymb,mathtools,makecell,amssymb,bbm, amsthm,dsfont,color,times,txfonts,nicefrac,framed,float,enumitem,tikz,physics,circuitikz}
\usepackage{algpseudocode}
\usepackage[symbol]{footmisc}
\usepackage{amsfonts}
\usepackage[normalem]{ulem}
\usepackage{tcolorbox}\tcbset{before skip=10pt,toptitle=1mm,bottomtitle=1mm,fonttitle=\bfseries}
\tcbuselibrary{theorems}
\tcbuselibrary{breakable}
\newtcolorbox[auto counter]{pabox}[2][]{%
	colback=white,colframe=black!10,fonttitle=\bfseries,coltitle=black,breakable,title=#2,#1}

\definecolor{equationcolor}{RGB}{222,94,100}

\usepackage[ruled,vlined]{algorithm2e}
\usepackage[mathscr]{eucal}
\usepackage{eucal}
\usepackage{ragged2e}

\SetKwComment{Comment}{/* }{ */}
\SetKwInOut{Input}{Input}
\SetKwInOut{Parameters}{Parameters}
\SetKwInOut{Initialization}{Initialization}
\SetKwInOut{Set}{Set}
\SetKwInOut{Output}{Output\,}

\newlength{\commentWidth}
\begin{document}
%
\title{Thermodynamic Algorithms for Quadratic Programming}


\author{\IEEEauthorblockN{Patryk-Lipka Bartosik\IEEEauthorrefmark{1}\IEEEauthorrefmark{2},
Kaelan Donatella\IEEEauthorrefmark{1}\IEEEauthorrefmark{4},
Maxwell Aifer\IEEEauthorrefmark{4}, 
Denis Melanson\IEEEauthorrefmark{4},\\ Marti Perarnau-Llobet\IEEEauthorrefmark{2}, Nicolas Brunner\IEEEauthorrefmark{2} and
Patrick J. Coles\IEEEauthorrefmark{4}}

\IEEEauthorblockA{\IEEEauthorrefmark{2}University of Geneva, Switzerland}
\IEEEauthorblockA{\IEEEauthorrefmark{4} Normal Computing, New York, NY, USA}
\IEEEauthorblockA{\IEEEauthorrefmark{1} These authors contributed equally.}
}

%


\maketitle


\begin{abstract}
Thermodynamic computing has emerged as a promising paradigm for accelerating computation by harnessing the thermalization properties of physical systems. This work introduces a novel approach to solving quadratic programming problems using thermodynamic hardware. By incorporating a thermodynamic subroutine for solving linear systems into the interior-point method, we present a hybrid digital-analog algorithm that outperforms traditional digital algorithms in terms of speed. Notably, we achieve a polynomial asymptotic speedup compared to conventional digital approaches. Additionally, we simulate the algorithm for a support vector machine and predict substantial practical speedups with only minimal degradation in solution quality. Finally, we detail how our method can be applied to portfolio optimization and the simulation of nonlinear resistive networks. 
\end{abstract}


%
\IEEEpeerreviewmaketitle

\section{Introduction}
Hard optimization problems are ubiquitous in machine learning, artificial intelligence, finance, engineering, and bioinformatics. Standard digital computers often encounter difficulties or inefficiencies in solving such optimization problems. For example, digital devices are inherently discrete, which can make it difficult to efficiently explore continuous solution spaces for optimization problems involving continuous variables. Moreover, the fact that digital operations require active energy injection, as opposed to passively allowing physical processes to occur, makes digital devices energy hungry.

This motivates exploring alternative computing paradigms for which optimization is more naturally suited. One possibility is employing devices where optimization naturally occurs, due to inherent physical processes that take place on the device. Ising machines~\cite{Mohseni2022,Bybee2023,Tanahashi2019,chou2019analog,yamamoto2020coherent,wang2019oim,patel2024pass}, quantum annealers~\cite{Hauke_2020} and probabilistic bit (p-bit) based systems\cite{Camsari_2019,kaiser2022life,chowdhury2023full,aadit2023all} have also been employed for this purpose in recent years, with particular focus on discrete optimization problems, which are well-suited to the binary nature of the hardware that is typically employed. Recently, hybrid continuous-discrete optimization has been considered with an analog computing device, specifically a hybrid optical-electrical system~\cite{kalinin2023analogiterativemachineaim}.

The rise of thermodynamic computing~\cite{conte2019thermodynamic, wiredarticle} motivates the use of such computers for optimization. Namely, it is natural to target optimization problems with thermodynamic computers, because such computers naturally relax to equilibrium, which can be viewed as minimizing the free energy (and hence solving some optimization problem). Previously, thermodynamic computers have been proposed for modeling neural networks~\cite{hylton2020thermodynamic,lipka2024thermodynamic,hylton2022thermodynamic}, second-order training of neural networks~\cite{donatella2024thermodynamic}, solving linear algebra problems~\cite{aifer2024npj,duffield2023thermodynamic}, generative artificial intelligence (AI)~\cite{coles2023thermodynamic}, and Bayesian inference~\cite{coles2023thermodynamic,aifer2024thermodynamic}. Small-scale experimental demonstrations~\cite{melanson2023thermodynamic} of Gaussian sampling and matrix inversion with CMOS-compatible hardware highlight the potential to scale thermodynamic computers with standard silicon fabrication methods. 



Given the suitability of thermodynamic systems for continuous optimization, in this paper we explore their application to quadratic programming, which involves minimizing quadratic objective function with linear constraints. Through careful consideration, we show that quadratic programming can be mapped onto thermodynamic computing hardware that was previously considered in the literature, in Refs.~\cite{aifer2024npj,aifer2024thermodynamic,aifer2024error} and~\cite{duffield2023thermodynamic}. Specifically, we present a thermodynamic algorithm for quadratic programming with the interior point method that achieves polynomial speedup compared to conventional methods. This algorithm involves offloading the linear system subroutine to a stochastic processing unit (SPU) while computing matrix-vector and matrix-matrix multiplications with a GPU. The SPU contains a crossbar-array encoding input matrices, similar to that presented in Ref.~\cite{sun2019solving}, along with capacitors for time-integration. This hybrid computation is similar to the strategy employed in Ref.~\cite{donatella2024thermodynamic}, where one leverages the different strengths of each compute block (GPU and SPU). Based on realistic hardware assumptions, we predict a $10-30\times$ speedup for $\sim 1000$ dimensions with respect to digital methods on a support vector machine task using a simulation of our thermodynamic hardware.



Our work opens up the possibility of fast, energy-efficient solutions of quadratic optimization problems driven by thermodynamic processes. We highlight three key applications of our thermodynamic algorithm: support vector machines in machine learning, portfolio optimization in finance and simulation of non-linear resistive networks in hardware design for AI applications. The latter represents an opportunity to use thermodynamic hardware for designing other unconventional hardware. Moreover, we briefly discuss the extension to non-quadratic problems through sequential quadratic programming, which would allow our proposed algorithm to be used for a wide range of optimization problems. 



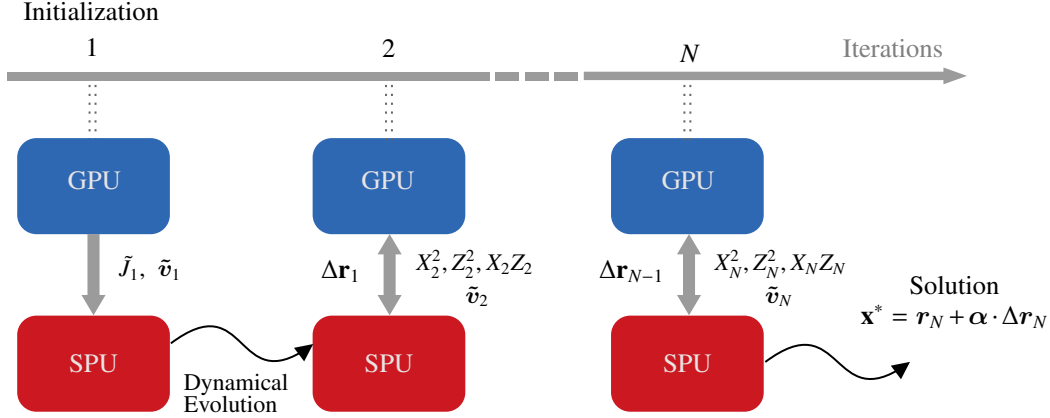
\begin{figure*}[t]%
\centering
\scalebox{1}{\tikzset{every picture/.style={line width=0.75pt}} 

\begin{tikzpicture}[x=0.75pt,y=0.75pt,yscale=-1,xscale=1]

\draw [color={rgb, 255:red, 104; green, 104; blue, 104 }  ,draw opacity=1 ] [dash pattern={on 0.84pt off 2.51pt}]  (135.5,75.59) -- (135.5,110.09)(132.5,75.59) -- (132.5,110.09) ;
\draw  [draw opacity=0][fill={rgb, 255:red, 204; green, 25; blue, 33 }  ,fill opacity=1 ] (96,206.2) .. controls (96,200.84) and (100.34,196.5) .. (105.7,196.5) -- (163.3,196.5) .. controls (168.66,196.5) and (173,200.84) .. (173,206.2) -- (173,235.3) .. controls (173,240.66) and (168.66,245) .. (163.3,245) -- (105.7,245) .. controls (100.34,245) and (96,240.66) .. (96,235.3) -- cycle ;
\draw  [draw opacity=0][fill={rgb, 255:red, 204; green, 25; blue, 33 }  ,fill opacity=1 ] (245,206.2) .. controls (245,200.84) and (249.34,196.5) .. (254.7,196.5) -- (312.3,196.5) .. controls (317.66,196.5) and (322,200.84) .. (322,206.2) -- (322,235.3) .. controls (322,240.66) and (317.66,245) .. (312.3,245) -- (254.7,245) .. controls (249.34,245) and (245,240.66) .. (245,235.3) -- cycle ;
\draw  [draw opacity=0][fill={rgb, 255:red, 204; green, 25; blue, 33 }  ,fill opacity=1 ] (395.5,206.2) .. controls (395.5,200.84) and (399.84,196.5) .. (405.2,196.5) -- (462.8,196.5) .. controls (468.16,196.5) and (472.5,200.84) .. (472.5,206.2) -- (472.5,235.3) .. controls (472.5,240.66) and (468.16,245) .. (462.8,245) -- (405.2,245) .. controls (399.84,245) and (395.5,240.66) .. (395.5,235.3) -- cycle ;
\draw  [draw opacity=0][fill={rgb, 255:red, 46; green, 104; blue, 179 }  ,fill opacity=1 ] (96,116.6) .. controls (96,111.24) and (100.34,106.9) .. (105.7,106.9) -- (163.3,106.9) .. controls (168.66,106.9) and (173,111.24) .. (173,116.6) -- (173,145.7) .. controls (173,151.06) and (168.66,155.4) .. (163.3,155.4) -- (105.7,155.4) .. controls (100.34,155.4) and (96,151.06) .. (96,145.7) -- cycle ;
\draw  [draw opacity=0][fill={rgb, 255:red, 46; green, 104; blue, 179 }  ,fill opacity=1 ] (245,116.6) .. controls (245,111.24) and (249.34,106.9) .. (254.7,106.9) -- (312.3,106.9) .. controls (317.66,106.9) and (322,111.24) .. (322,116.6) -- (322,145.7) .. controls (322,151.06) and (317.66,155.4) .. (312.3,155.4) -- (254.7,155.4) .. controls (249.34,155.4) and (245,151.06) .. (245,145.7) -- cycle ;
\draw  [draw opacity=0][fill={rgb, 255:red, 46; green, 104; blue, 179 }  ,fill opacity=1 ] (395.5,116.6) .. controls (395.5,111.24) and (399.84,106.9) .. (405.2,106.9) -- (462.8,106.9) .. controls (468.16,106.9) and (472.5,111.24) .. (472.5,116.6) -- (472.5,145.7) .. controls (472.5,151.06) and (468.16,155.4) .. (462.8,155.4) -- (405.2,155.4) .. controls (399.84,155.4) and (395.5,151.06) .. (395.5,145.7) -- cycle ;
\draw    (173.5,210.5) .. controls (212.7,181.1) and (205.61,238.46) .. (242.96,212.53) ;
\draw [shift={(245.3,210.83)}, rotate = 143.13] [fill={rgb, 255:red, 0; green, 0; blue, 0 }  ][line width=0.08]  [draw opacity=0] (8.93,-4.29) -- (0,0) -- (8.93,4.29) -- cycle    ;

\draw    (473.3,219.9) .. controls (512.5,190.5) and (506.56,247.47) .. (543.96,221.53) ;
\draw [shift={(546.3,219.83)}, rotate = 143.13] [fill={rgb, 255:red, 0; green, 0; blue, 0 }  ][line width=0.08]  [draw opacity=0] (8.93,-4.29) -- (0,0) -- (8.93,4.29) -- cycle    ;
\draw [color={rgb, 255:red, 104; green, 104; blue, 104 }  ,draw opacity=1 ] [dash pattern={on 0.84pt off 2.51pt}]  (285,75.59) -- (285,110.09)(282,75.59) -- (282,110.09) ;
\draw [color={rgb, 255:red, 104; green, 104; blue, 104 }  ,draw opacity=1 ] [dash pattern={on 0.84pt off 2.51pt}]  (435.5,75.59) -- (435.5,110.09)(432.5,75.59) -- (432.5,110.09) ;
\draw  [draw opacity=0][fill={rgb, 255:red, 155; green, 155; blue, 155 }  ,fill opacity=1 ] (137.53,155.33) -- 
(137.53,165.58) -- (137.53,186.08) -- (140.57,186.08) -- (134.49,196.33) -- (128.42,186.08) -- (131.45,186.08) -- 
(131.45,165.58) --
(131.45,155.33) --
cycle ;
\draw  [draw opacity=0][fill={rgb, 255:red, 155; green, 155; blue, 155 }  ,fill opacity=1 ] (283.83,155.33) -- (289.9,165.58) -- (286.86,165.58) -- (286.86,186.08) -- (289.9,186.08) -- (283.83,196.33) -- (277.75,186.08) -- (280.79,186.08) -- (280.79,165.58) -- (277.75,165.58) -- cycle ;
\draw  [draw opacity=0][fill={rgb, 255:red, 155; green, 155; blue, 155 }  ,fill opacity=1 ] (434.83,155.33) -- (440.9,165.58) -- (437.86,165.58) -- (437.86,186.08) -- (440.9,186.08) -- (434.83,196.33) -- (428.75,186.08) -- (431.79,186.08) -- (431.79,165.58) -- (428.75,165.58) -- cycle ;
\draw  [draw opacity=0][fill={rgb, 255:red, 155; green, 155; blue, 155 }  ,fill opacity=1 ] (90.75,73.5) -- (334.25,73.5) -- (334.25,77.88) -- (90.75,77.88) -- cycle ;
\draw  [draw opacity=0][fill={rgb, 255:red, 155; green, 155; blue, 155 }  ,fill opacity=1 ] (337.25,73.5) -- (349.25,73.5) -- (349.25,78) -- (337.25,78) -- cycle ;
\draw  [draw opacity=0][fill={rgb, 255:red, 155; green, 155; blue, 155 }  ,fill opacity=1 ] (352.25,73.5) -- (364.25,73.5) -- (364.25,78) -- (352.25,78) -- cycle ;
\draw  [draw opacity=0][fill={rgb, 255:red, 155; green, 155; blue, 155 }  ,fill opacity=1 ] (366.75,73.5) -- (378.75,73.5) -- (378.75,78) -- (366.75,78) -- cycle ;
\draw  [draw opacity=0][fill={rgb, 255:red, 155; green, 155; blue, 155 }  ,fill opacity=1 ] (381.75,73.38) -- (563.15,73.38) -- (563.15,71.25) -- (575,75.5) -- (563.15,79.75) -- (563.15,77.63) -- (381.75,77.63) -- cycle ;

\draw (510.97,55.53) node [anchor=north west][inner sep=0.75pt]  [color={rgb, 255:red, 128; green, 128; blue, 128 }  ,opacity=1 ] [align=left] {Iterations};
\draw (129,55.17) node [anchor=north west][inner sep=0.75pt]   [align=left] {$\displaystyle 1$};
\draw (278.5,55.5) node [anchor=north west][inner sep=0.75pt]   [align=left] {$\displaystyle 2$};
\draw (428.65,58) node [anchor=north west][inner sep=0.75pt]   [align=left] {$\displaystyle N$};
\draw (134.5,220.75) node  [color={rgb, 255:red, 232; green, 228; blue, 228 }  ,opacity=1 ] [align=left] {\begin{minipage}[lt]{43.86pt}\setlength\topsep{0pt}
\begin{center}
SPU
\end{center}

\end{minipage}};
\draw (134.95,128.3) node  [color={rgb, 255:red, 232; green, 228; blue, 228 }  ,opacity=1 ] [align=left] {\begin{minipage}[lt]{52.97pt}\setlength\topsep{0pt}
\begin{center}
GPU
\end{center}

\end{minipage}};
\draw (178.5,222) node [anchor=north west][inner sep=0.75pt]  [font=\scriptsize] [align=left] {\begin{minipage}[lt]{32.72pt}\setlength\topsep{0pt}
\begin{center}
{\small Dynamical}\\{\small Evolution}
\end{center}

\end{minipage}};

\draw (145.33,165) node [anchor=north west][inner sep=0.75pt]  [font=\small]  {$\tilde{J}_{1} ,\ \bm{\tilde{v}}_{1}$};

\draw (520,175) node [anchor=north west][inner sep=0.75pt]   [align=left] {\begin{minipage}[lt]{70.87pt}\setlength\topsep{0pt}
\begin{center}
Solution \\$\displaystyle \mathbf{x}^{*} = \bm{r}_N + \bm{\alpha} \cdot \Delta \bm{r}_N$
\end{center}

\end{minipage}};
\draw (283.05,128.3) node  [color={rgb, 255:red, 232; green, 228; blue, 228 }  ,opacity=1 ] [align=left] {\begin{minipage}[lt]{52.97pt}\setlength\topsep{0pt}
\begin{center}
GPU
\end{center}

\end{minipage}};
\draw (434.45,128.3) node  [color={rgb, 255:red, 232; green, 228; blue, 228 }  ,opacity=1 ] [align=left] {\begin{minipage}[lt]{52.97pt}\setlength\topsep{0pt}
\begin{center}
GPU
\end{center}

\end{minipage}};
\draw (283.5,220.75) node  [color={rgb, 255:red, 232; green, 228; blue, 228 }  ,opacity=1 ] [align=left] {\begin{minipage}[lt]{43.86pt}\setlength\topsep{0pt}
\begin{center}
SPU
\end{center}

\end{minipage}};
\draw (434,220.75) node  [color={rgb, 255:red, 232; green, 228; blue, 228 }  ,opacity=1 ] [align=left] {\begin{minipage}[lt]{43.86pt}\setlength\topsep{0pt}
\begin{center}
SPU
\end{center}

\end{minipage}};
\draw (295.57,162) node [anchor=north west][inner sep=0.75pt]  [font=\small]  {$X_{2}^2, Z_{2}^2 ,X_{2} Z_{2}$};
\draw (320.57,180.23) node [anchor=north west][inner sep=0.75pt]  [font=\small]  {$\bm{\tilde{v}}_{2}$};
\draw (388,167)node [anchor=north west][inner sep=0.75pt]    {$\Delta \mathbf{r}_{N-1}$};
\draw  (446,162) node [anchor=north west][inner sep=0.75pt]  [font=\small]  {$X_{N}^2, Z_{N}^2 ,X_{N} Z_{N}$};
\draw (471.67,180.23) node [anchor=north west][inner sep=0.75pt]  [font=\small]  {$\bm{\tilde{v}}_{N}$};
\draw (94.5,36.6) node [anchor=north west][inner sep=0.75pt]  [font=\normalsize] [align=left] {\begin{minipage}[lt]{56.02pt}\setlength\topsep{0pt}
\begin{center}
Initialization
\end{center}

\end{minipage}};
 
\draw (247.83,167) node [anchor=north west][inner sep=0.75pt]    {$\Delta \mathbf{r}_{1}$};

\end{tikzpicture}}
\caption{\justifying \textbf{Overview of the hybrid digital-analog algorithm to solve quadratic programs.} At initialization, $\tilde{J}_1 = J_1^\top J_1$ and  $v_1$ are computed by a GPU (to benefit from the high parallelization of matrix-matrix and matrix-vector multiplications) and uploaded onto the SPU. After some dynamical evolution, the update $\Delta \bm{r}_1$ is downloaded from the SPU, which enables the calculation of quantities $X_2, Z_2, X_2Z_2$ which are used to upload the matrix representation of $\tilde{J}$ (only updating blocks of it), and $v_2$ which are then downloaded onto the SPU. This continues until a satisfactory solution is reached, according to the criteria described in~\ref{alg:2}. }\label{fig:Flowchart}
\end{figure*}
\setcounter{section}{1}
\section{Background}
\subsection{Quadratic Programming}
\label{sec:qprog}
Quadratic programs are convex optimization problems of the following form:
\begin{align}
    \label{eq:qprog}
    \min_{\bm{x} \in \mathbb{R}^n}& \quad V(\bm{x}) := \frac{1}{2} \bm{x}^\top Q \bm{x} + \bm{c}^\top \bm{x}  \\
    \text{subject to}& \quad    A \bm{x} = \bm{b}, \ \nonumber\\
    &\quad \bm{x} \geq 0. \nonumber
\end{align}
where $\bm{x}^\top$ denotes transpose of the vector $\bm{x}$ and $Q \in \mathbb{R}^{n \times n}$ is a positive semidefinite matrix. In what follows, we consider the case when $A \in \mathbb{R}^{m \times n}$ (with $m \leq n$) has full rank, $\bm{b}\in \mathbb{R}^{m}$ and $\bm{c}\in \mathbb{R}^{n}$. Unless otherwise stated, all vectors and matrices considered in this work are real. The optimal solution to the problem \eqref{eq:qprog} will be denoted with $\bm{x}^{\star} := \text{argmin} V(\bm{x})$ (under constraints). The size of the problem is specified by two numbers: $n$ (dimension of the optimization variable $\bm{x}$) and $m$ (number of equality constraints).

Note that linear programming \cite{vanderbei1998linear}, which has applications in operations research including maximizing profits and minimizing costs~\cite{SINUANYSTERN20231069}, constitutes a subset of this framework and corresponds to taking $Q = 0$.  
More broadly, quadratic programming has applications in machine learning (e.g., support vector machines), portfolio optimization in finance, and hardware design. We feature some of these applications in Sec.~\ref{sec:app}.

\subsection{Interior Point Methods}
\label{sec:IPMs}
Interior-point methods (IPMs) constitute a widely used class of algorithms for solving quadratic programs \cite{GONDZIO2012587}. The basic idea of IPMs is to iteratively construct a sequence of suboptimal solutions (known as \emph{central path}) that converges to the optimal solution ~$\bm{x}^\star$. The starting point for an IPM algorithm is to transform the optimization problem \eqref{eq:qprog} into a set of (potentially nonlinear) equations using the Lagrangian duality theory \cite{boyd2004convex} involving primal $\bm{x}$, dual $\bm{y}$, as well as slack variables $\bm{z}$. These equations are then solved iteratively using Newton's method \cite{hamming2012numerical}. Namely, starting from an initial point $\bm{r}_0 = (\bm{x}_0, \bm{y}_0, \bm{z}_0)$ one generates successive iterates given by
\begin{align}   \label{eq:central_path}
    \bm{r}_{k+1} = \bm{r}_k + \alpha (\Delta \bm{x}_k, \Delta \bm{y}_k, \Delta \bm{z}_k),
\end{align}
where $\Delta \bm{x}_k$, $\Delta\bm{y}_k$ and $\Delta \bm{z}_k$ specify the direction of step $k$ in the space involving primal, dual and slack variables. The scalar $\alpha$ specifies the step size and is usually selected so that $\bm{x}_{k+1} > 0$ and $\bm{z}_{k+1} > 0$. The direction of the step is obtained by solving a system of linear equations
\begin{align}
    \label{eq:lin_sys}
    J_k \begin{bmatrix}
        \Delta \bm{x}_k \\
        \Delta \bm{y}_k \\
        \Delta \bm{z}_k
    \end{bmatrix} = \bm{v}_k, 
\end{align}
where $J_k \in \mathbb{R}^{(2n+m) \times (2n+m)}$ is the Jacobian associated with the system of Karush–Kuhn–Tucker (KKT) conditions and $\bm{v}_k \in \mathbb{R}^{2n + m}$ is a vector capturing both fixed parameters of the initial problem as well as tunable parameters of the central path. These quantities are defined as:
\begin{align}
   J_k &= \begin{bmatrix}
    \label{eq:ipm_lin}
        -Q & A^{\top} & I \\
        A & 0 &0 \\
        Z_k &0 &X_k
    \end{bmatrix}, \\ \label{eq:ipm_vec} \bm{v}_k &=
    \begin{bmatrix}
        Q \bm{x}_k - A^{\top} \bm{y}_k - \bm{z}_k + \bm{c} \\
        -A\bm{x}_k + \bm{b} \\
        -X_k Z_k \bm{e} + \mu_k \sigma_k \bm{e},
    \end{bmatrix}
\end{align}
where $I$ is the identity matrix, $\sigma_{k} \in (0, 1)$ is a barrier parameter and $\bm{e}$ is the identity vector $(1, \ldots, 1)$. Moreover, for clarity of notation we introduced diagonal matrices $X_k = \text{diag}(x_1, \ldots, x_n), Z_k = \text{diag}(z_1, \ldots, z_n) $. By iteratively solving Eq. \eqref{eq:lin_sys} one obtains the \emph{central path} $\{\bm{r}_k \in \mathbb{R}^{2n+m}\,|\, k = 0, 1, \ldots, N\}$, from which one can determine an $\epsilon$-accurate solution of the problem \eqref{eq:qprog} in 
\begin{align}\label{eq:bound_iters}
    N = \mathcal{O}\left(\sqrt{(n+m)} \log(1/\epsilon)\right)
\end{align}iterations \cite{GONDZIO2012587}. The most expensive task during an interior point iteration is finding the solution of the linear system of equations \eqref{eq:lin_sys}. At the same time, since the system from Eq. \eqref{eq:lin_sys} is only an approximation to the original set of KKT conditions, interior point algorithms do not require solving it to a high accuracy.  These two facts make interior point algorithms good candidates for hardware acceleration of linear systems with approximate solutions given by thermodynamic computation. In what follows we discuss how to implement an interior point algorithm using a thermodynamic algorithm for the linear system solver\cite{aifer2024npj}. More details on the interior point algorithm used in this work are provided in Appendix \ref{sec:IPMs}.

\subsection{Thermodynamic Linear Solver}
\label{sec:thermo_comp}
Thermodynamic algorithms attempt to solve problems by leveraging the natural tendency of physical systems to reach thermal equilibrium. In this work we use a recent result that provides an efficient thermodynamic algorithm for solving linear systems~\cite{aifer2024npj}. To understand this key subroutine let us consider a classical system specified by its positions $\bm{x} = (x_1, \ldots)$ and momenta $\bm{p} = (p_1, \ldots)$, and governed by a quadratic Hamiltonian
\begin{align}
    \label{eq:quad_hamiltonian}
    \mathcal{H}(\bm{x}, \bm{p}) = \frac{1}{2m} \bm{p}^\top \bm{p} + V(\bm{x}),
\end{align}
where $V(\bm{x}):=\frac{1}{2} \bm{x}^{\top} Q \bm{x} - \bm{c}^{\top} \bm{x}$ is the potential function. This Hamiltonian captures a broad range of physical systems, e.g. a network of coupled mechanical oscillators or an electric circuit composed of resistors, inductors and capacitors. Notably, when the system described by Eq. \eqref{eq:quad_hamiltonian} is coupled with a thermal environment at temperature $T$ and allowed to equilibrate, it reaches a thermal state described by the Boltzman distribution $\pi(\bm{x}, \bm{p}) \propto e^{-\mathcal{H}(\bm{x},\bm{p})/T}$. This state, after marginalizing to $\bm{x}$, yields $\pi(\bm{x}) = \mathcal{N}(Q^{-1} \bm{c}, T Q^{-1})$, where $\mathcal{N}(\mu, \sigma)$ is a Gaussian with mean $\mu$ and variance $\sigma$. 

For our purposes it is important to observe that at thermal equilibrium the first moment $\langle \bm{x} \rangle$ is given by $\langle \bm{x} \rangle = Q^{-1} \bm{c}$. This value can be arbitrarily well approximated by calculating the time average
\begin{align}
    \label{eq:approx_mean}
    \langle \bm{x} \rangle \approx \frac{1}{\tau} \int_{t_0}^{t_0 + \tau} \bm{x}(t)\,\text{d}t.
\end{align}
The fact that the average converges to the true mean of the thermal distribution (i.e., the approximation in Eq.~\eqref{eq:approx_mean} becomes better as $\tau$ grows) is guaranteed by the ergodic hypothesis \cite{boltzmann1871einige}. In this way we can essentially solve the system of equations $Q \bm{x} = c$, as originally presented in Ref. \cite{aifer2024npj}. This reasoning is restated as Algorithm \ref{alg:1} for the convenience of the reader. In particular, the algorithm requires choosing two equilibration tolerance parameters, $\epsilon_{\mu}$ and $\epsilon_{\Sigma}$. These specify the burn-in time $t_0 = t_0(\epsilon_{\mu}, \epsilon_{\Sigma})$, which is the time before samples are collected. Furthermore, the algorithm requires specifying an integration time $\tau = \tau(\epsilon_x, p_{\text{succ}})$ that depends on the error tolerance $\epsilon_x$ specifying the approximation of the integral from Eq. \eqref{eq:approx_mean}, as well as the probability of success $p_{\text{succ}}$. For a more detailed analysis of the algorithm and the role of input parameters, see Ref. \cite{aifer2024npj}.
    
\begin{algorithm}[h!]
\caption{Thermodynamic Linear Solver (TLS)}\label{alg:1}
\Input{\hspace{1pt} Matrix $Q \in \mathbb{R}^{n \times n}$ \\ \hspace{0.5pt} Vector $c \in \mathbb{R}^n$ \\ \hspace{0.5pt} Error tolerance $\epsilon_{\mu}, \epsilon_{\Sigma}, \epsilon_{x} > 0$, \\ \hspace{0.5pt} Success probability $p_{\text{succ}} > 0$.  }
\Output{The approximate solution $x \in \mathbb{R}^n$ to $Q \bm{x} = \bm{c}$}
\vspace{5pt}
\begin{enumerate}
    \item Set the potential of the physical system to $V(\bm{x}) = \frac{1}{2}\bm{x}^{\top} Q \bm{x} - \bm{c}^{\top} \bm{x}$.
    \item Evolve the system for time $t_0 = t_0(\epsilon_{\mu}, \epsilon_{\Sigma})$ to reach thermal equilibrium.
    \item Measure the average $\bar{\bm{x}}$ using $\tau = \tau(\epsilon_x, p_{\text{succ}})$, i.e.
    \begin{align}
        \bar{\bm{x}} = \frac{1}{\tau} \int_{t_0}^{t_0 + \tau} \bm{x}(t) \,\text{d}t.\nonumber
    \end{align}
\end{enumerate} 
\textbf{Return} $\bar{x}$   
\end{algorithm}

\section{Thermodynamic algorithms for solving quadratic programs}
\label{sec:thermo_quad}
We now present several approaches for solving quadratic optimization problems \eqref{eq:qprog} using a thermodynamic linear solver.

\begin{table*}[ht!]
\small
 \caption{\small Asymptotic complexities of solving quadratic programs with interior point methods.}
    \centering
    \begin{tabular}{c|c}
    Method & Runtime \\ \hline
   
   LU decomposition & $\mathcal{O}\left((n+m)^{7/2}\log(1/\epsilon)\right)$ \\
        Conjugate gradients & $\mathcal{O}\left(n^3 + m^3 + \sqrt{n+m}\log(1/\epsilon)[n^2 + m^2 +\sqrt{\kappa}(n+m)^2]\right)$\\
        Thermodynamic & $\mathcal{O}\left(n^3 + m^3 +\sqrt{n+m}\log(1/\epsilon)[n^2 + m^2 + \kappa^2(n+m)\delta^{-2}]\right)$ 
    \end{tabular}
   
    \label{tab:complexities}
\end{table*}
\subsection{Unconstrained and equality constrained QP}
The simplest instance of a quadratic program corresponds to the case when there are no constraints in Eq. \eqref{eq:qprog}, i.e. when $A = 0$. In this case the problem can be solved analytically. That is, the optimum  $\bm{x}^{\star}$ satisfies $\nabla V(\bm{x}^{\star}) = 0$, i.e.
\begin{align}
    \label{eq:xstar_unc}
    \bm{x}^{\star} = Q^{-1} \bm{c}.
\end{align}
Hence solving the quadratic program in this case requires $(i)$ inverting the matrix $Q$ and $(ii)$ performing matrix-vector multiplications. Matrix inversion generally requires $\mathcal{O}(n^3)$ operations on a conventional digital computer. On the other hand, these two operations can be performed simultaneously using a thermodynamic computer. More specifically, by encoding the matrix $Q$ and vector $\bm{c}$ in the Hamiltonian of the system as done in Eq. \eqref{eq:quad_hamiltonian}, the optimum $\bm{x}^{\star}$ can be approximated by calculating the time average $\langle \bm{x} \rangle$ from Eq. \eqref{eq:approx_mean}. In Ref.~\cite{aifer2024npj}, it is shown that solving a linear system with error $\delta$ has runtime complexity $\mathcal{O}(n\kappa^2\delta^{-2})$, with $\kappa$ the condition number of the matrix entering the linear system.

It is worth mentioning that a similar approach as outlined above can be used to solve quadratic programs with linear {equality} constraints. In order to see this, observe that such linear constraints can be incorporated by transforming the problem into an unconstrained case using the method of \emph{Lagrange multipliers}. More specifically, consider the problem from Eq. \eqref{eq:qprog} with an equality constraint $A \bm{x} = \bm{b}$. By introducing a new variable $\bm{\mu} \in \mathbb{R}^n$ we can form the Lagrangian
\begin{align}
    L(\bm{x}, \bm{\mu}) &= V(\bm{x}) + \bm{\mu}^{\top}(A\bm{x} - \bm{b}) \\
    &= \frac{1}{2} \bm{x}^{\top} Q \bm{x} + (A^{\top} \mu - \bm{c})^{\top} x - \mu^{\top} \bm{c}.
\end{align}

The stationary point of the Lagrangian corresponds to the optimum $\bm{x}^{\star}$ of our original problem. It can be determined by setting the partial derivatives of $L(\bm{x}, \bm{\mu})$ to zero, i.e. 
\begin{align}
    \frac{\partial L(\bm{x},\bm{\mu})}{\partial \bm{x}} &= 0 \implies Q \bm{x} + A^\top \bm{\mu} = \bm{c}, \\
    \frac{\partial L(\bm{x},\bm{\mu})}{\partial \bm{\lambda}} &= 0 \implies A \bm{x} = \bm{b}. 
\end{align}
The unique global optimum $(\bm{x}^{\star}, \bm{\mu}^{\star})$ can be found by solving the linear system of equations  
\begin{align}
    \begin{pmatrix}
Q & A^\top \\
A & 0
\end{pmatrix} 
\begin{pmatrix}
\bm{x}  \\
\bm{\mu}
\end{pmatrix} = 
\begin{pmatrix}
\bm{c}  \\
\bm{b}
\end{pmatrix} \implies \begin{pmatrix}
\bm{x}^{\star} \\
\bm{\lambda}^{\star}
\end{pmatrix} =  \begin{pmatrix}
Q & A^\top \\
A & 0
\end{pmatrix}^{-1} \begin{pmatrix}
\bm{c}  \\
\bm{b}
\end{pmatrix}.
\end{align}
Notice that this linear system has exactly the same form as Eq. \eqref{eq:xstar_unc}, the only difference being that the dimension of the variables is now twice as large. 

\subsection{Inequality constrained QPs} 
In general, quadratic programs as displayed in Eq. \eqref{eq:qprog} may contain inequality constraints. This form of QPs also encompasses both unconstrained and equality constrained QPs that we discussed in the previous section. One potential way to address them is to use the interior-point algorithm as outlined in Sec. \ref{sec:qprog}. In this type of algorithm, the most computationally expensive step is to calculate the direction vector $\Delta \bm{r}$ of the Newton step [see Eq. \eqref{eq:lin_sys}]. This type of algorithm, also known in the literature as the short-step path-following algorithm, is guaranteed to converge to an $\epsilon$-accurate solution in $\mathcal{O}(\sqrt{(n+m)} \log(1/\epsilon))$ iterations of Newton step  \cite{GONDZIO2012587}.

In the original interior-point algorithm each iteration requires solving a (potentially sparse) linear system of equations. Notably, our observation here is that such a linear system can be approximately solved more efficiently using the thermodynamic subroutine described in Sec. \ref{sec:thermo_comp}. This leads to a variant of an interior-point algorithm where the most computationally demanding step is handled by an efficient thermodynamic solver (TLS). The algorithm \ref{alg:2} outlines this procedure which finds the approximate solution to the quadratic program in Eq. \eqref{eq:qprog}. In the next sections, we investigate the algorithm in more detail, and discuss the technical modifications necessary to adapt the thermodynamic solver (TLS) for use within the interior-point framework.

\subsubsection{Linear systems from the normal equations}
The interior point algorithm generates points on the central path $\bm{r}_k = (\bm{x}_k, \bm{y}_k, \bm{z}_k)$ as specified by Eq. \eqref{eq:central_path} 
In order to determine these points one has to solve the system of linear equations given in Eq. \eqref{eq:linearized} for each $k \in \{1, \ldots, N\}$, where $N$ is the total number of iterations. If one wishes to use conjugate gradients or the thermodynamic hardware, the matrix $A$ in the linear system $Ax=b$ one wishes to solve has to be made positive.

Note that the linear system corresponding to Eq.~\eqref{eq:ipm_lin} may be symmetrized by constructing $J_k ^\top J_k$, given by:
\begin{align}\label{eq:jdagj}
    J_k ^\top J_k = 
     \begin{pmatrix}
    Q^\top Q + A^\top A + Z_k^2 & -Q^{\top} A^{\top} & X_k Z_k - Q^{\top}  \\
    - A Q & A A^{\top} & A \\
    -Q + X_k Z_k & A^{\top} & I + X_k^2
\end{pmatrix}    .
\end{align}
The elements involving $Q, A$ and their transposes can be pre-computed at initialization, thus avoiding the need for matrix-matrix multiplications during each step of the IPM. We may then solve the linear system defined as 
\begin{align}
    J_k^\top J_k \bm{\Delta r}_k = J_k \bm{v}_k \iff
    \tilde{J}_k \bm{\Delta r}_k = \tilde{\bm{v}}_k,
\end{align} 
Crucially, only the blocks involving $X_k$ and $Z_k$, that is $X_k^2$, $Z_k^2$ and $X_k Z_k$, must be recomputed at each new iteration. The relevant blocks of $\tilde{J}_k$ may then be updated as, e.g. for the upper-left block:
\begin{equation}
    \tilde{J}_{k, [1:n, 1:n]} = \tilde{J}_{k-1, [1:n, 1:n]} + Z_k^2 - Z_{k-1}^2.
\end{equation}
which is an $\mathcal{O}(n)$ 
computation, as it only modifies the diagonal of the upper-left block and incurs the transfer of $n$ numbers to the SPU. Similar operations may be performed for all blocks that require updating, hence costing at most matrix-vector multiplications at each step. 

\begin{algorithm*}[h!]
\caption{Thermodynamic Interior Point Algorithm for Quadratic Programming}\label{alg:2}
\Parameters{
$\alpha_0 \in (0,1)$ stepsize factor;\\
\hspace{2pt} $\sigma \in (0, 1)$ barrier reduction parameter;\\
\hspace{2pt} $\epsilon_{p}, \epsilon_{d}, \epsilon_{o} \in (0,1)$ tolerance parameters; 

}
\Output{$\bm{x}^{\star}$ candidate for the optimal solution }
{\ }  \\

\Initialization{
$k = 0$ iteration counter; \\
\hspace{2pt} $\bm{r}_0 := (\bm{x}_0,\bm{y}_0,\bm{z}_0)$ initial point; \\
\hspace{2pt} $\mu_0 := \bm{x}_0^{\top} \bm{z}_0/n$ initial barrier parameter; \\
\hspace{2pt} $\bm{\xi}_p^0 := \bm{b} - A \bm{x}_{0}$ primal infeasibility; \\
\hspace{2pt} $\bm{\xi}_d^0 = \bm{c} - A^{\top} \bm{y}_0 - \bm{z}_0 + Q \bm{x}_0$ dual infeasibility; \\
\hspace{2pt} $J_k \leftarrow (\bm{r}_0, \mu_0, \sigma_0)$ via Eq. \eqref{eq:ipm_lin}; \\
\hspace{2pt} $\tilde{J}_k \leftarrow J_kJ_k$ via Eq.~\ref{eq:jdagj};\\
\hspace{2pt} \textit{condition} $=$ true; \\
}

\While{ condition }{
\hspace{3pt} $\mu_{k+1} = \sigma \mu_k$ {update barrier parameter};

\hspace{3pt} $J_k \leftarrow$ update matrix with $X_k, Z_k$; 

\hspace{3pt} $\bm{v}_k \leftarrow 
$ update via Eq. \eqref{eq:ipm_vec};

\hspace{3pt} $\bm{\tilde{v}}_k \leftarrow  J_k\bm{v}_k$;

\hspace{3pt} $\tilde{J}_k \leftarrow$ update matrix with $X_k^2, Z_k^2, X_kZ_k$; 

{\, }\\ 

\hspace{3pt} $\Delta \bm{r} \leftarrow \texttt{TLS}(\tilde{J}_k, \bm{\tilde{v}}_k)$ solve KKT system via Algorithm \ref{alg:1};

{\, }\\ 

\hspace{3pt} $\alpha_p = \alpha_0 \cdot \max \{\alpha \,:\, \bm{x}_k + \alpha \Delta \bm{x} \geq 0\}$ find primal stepsize; \\

\hspace{3pt} $\alpha_d = \alpha_0 \cdot \max \{\alpha \,:\, \bm{z}_k + \alpha \Delta \bm{z} \geq 0\}$ find dual stepsize; \\

\hspace{3pt} $\bm{r}_{k+1} \leftarrow \bm{r}_{k} + (\alpha_p \Delta \bm{x}, \alpha_d \Delta \bm{y}, \alpha_d \Delta \bm{x})$ make step; \\

{\, }\\ 

\hspace{3pt} $\bm{\xi}_p^{k+1} = \bm{b} - A \bm{x}_{k+1}$; \\

\hspace{3pt} $\bm{\xi}_d^{k+1} = \bm{c} - A^{\top} \bm{y}_{k+1} - \bm{z}_{k+1} + Q \bm{x}_{k+1}$; \\

{\, }\\ 

\hspace{3pt} \textit{condition} = $\left(\frac{\norm{\bm{\xi}_p^k}x}{1+\norm{\bm{b}}} > \epsilon_p\right)$ \textbf{or} $\left(\frac{\norm{\bm{\xi}_d^k}}{1+\norm{\bm{c}}} > \epsilon_d\right)$ \textbf{or} $\left(\frac{ \bm{x}_k^{\top} \bm{z}_k/n}{1+|V(\bm{x}_k)|} > \epsilon_o\right)$ with $V(\bm{x})$ given by Eq. \eqref{eq:qprog}.
}
\vspace{2pt}
\textbf{return}  $\bm{x}^{\star} := $ first $n$ elements of $\bm{r}_{k+1}$.
\end{algorithm*}

The above reduction in computational complexity is crucial for achieving a thermodynamic advantage in solving quadratic programs. In principle, one could imagine applying this technique to more general convex optimization problems by iteratively solving the corresponding KKT system. However, the matrix-matrix multiplication required for symmetrization in each iteration can generally negate any potential thermodynamic speedup. The ability to efficiently update the symmetric KKT system (involving $\tilde{J}_k$) for quadratic programs is therefore essential to the claims made in this paper.

\subsubsection{Algorithm}
This results in a thermodynamic algorithm for solving inequality constrained QPs detailed in Alg.~\ref{alg:2}, and pictorially represented in Fig.~\ref{fig:Flowchart}. 
At initialization, $\tilde{J}_k$ and $\bm{\tilde{v}}_k$ are computed and uploaded to the SPU. After some dynamical evolution time, the solution of the first linear system $\Delta \bm{r}_1$ is downloaded to the GPU, and the update to the trajectory $\bm{r}$ is performed. One then calculates the quantities $X_2, Z_2, X_2Z_2$ (all diagonal matrices) and $\bm{\tilde{v}}_2$ (which involves matrix-vector multiplications). This enables one to update $\tilde{J}$ efficiently directly on the SPU by only modifying diagonals of relevant blocks, thus avoiding $\mathcal{O}(n^3+m^3)$ costs.
The algorithm’s efficiency stems from offloading the sequential linear system solves to a thermodynamic computing platform, which has better asymptotic scaling—and can be extremely fast in practice due to the physical time constants of analog electronic systems—compared to digital alternatives. This operation is the only one that cannot be fully parallelized on a GPU, which limits the GPU’s use to its best-suited tasks. Additionally, constructing $\tilde{J}_k$ only needs to be performed during initialization using matrix multiplication, which is highly parallelizable on a GPU. Afterward, updates require only $\mathcal{O}(n)$ and $\mathcal{O}(m)$ operations per iteration, making the overall update and transfer from GPU to SPU efficient.
This algorithm leads to approximate updates $\bm{\Delta r}_k + \delta_k$, where $\delta_k$, the error in the linear system solution, is Gaussian with zero mean provided the steady-state of the thermodynamic system has been reached. The question of bounding the approximate solutions of the successive linear systems to solved has been studied in Refs~\cite{gondzio2013convergence} and~\cite{mizuno1999global}, and various update schemes on $\mu_k$ and $\alpha_d, \alpha_p$ are derived such that the additional error $\delta$ preserves the conditions for which the bound of Eq.~\ref{eq:bound_iters} is preserved (namely, that the iterates $\bm{\Delta r}_k$ remain in the $N_2$ neighbourhood of the path $\bm{r}_k$, see Ref.~\cite{GONDZIO2012587}).

\subsubsection{Computational complexity}

In Table~\ref{tab:complexities}, the complexities of the three considered algorithms for solving QPs with interior point algorithms are presented. LU decomposition, an exact algorithm and the most widely used for this task, involves $\mathcal{O}((n+m)^3)$ operations at each iteration. As mentioned previously, to reach error $\epsilon$ in the solution of the QP, the algorithm must run in $\sqrt{n+m} \log(1/\epsilon)$ iterations, thus resulting in the complexity reported in the table. The conjugate gradient algorithm may also be used, which involves similar steps to our thermodynamic algorithm; it requires building the $\tilde{J}_0$ matrix at initialization, incurring a $\mathcal{O}(n^3 + m^3)$, as well as matrix-vector multiplications at each iteration, costing $\mathcal{O}(n^2+m^2)$. Finally, the linear system solve using CG is $\mathcal{O}(\sqrt{\kappa}(n+m)^2)$, with $\kappa$ the condition number of $\tilde{J}_k$ (assuming it does not vary over iterations for simplicity).
The thermodynamic algorithm is very similar to CG, except that it solves a linear system in $\mathcal{O}(\kappa^2(n+m))$, resulting in the complexity shown in Table~\ref{tab:complexities}.
We also note that the the matrix-matrix and matrix-vector multiplications (the $n^3, n^2, m^3, m^2$ costs in the table for CG and the thermodynamic algorithm) are in practice extremely fast, as they are greatly parallelized on GPUs, which is not reflected in the asymptotic complexities. In the next section we show the practical speed of the algorithms employing LU and CG as well as the predicted advantage coming from a QP solved through a thermodynamic algorithm.

\section{Applications}
\label{sec:app}
In this section we discuss various applications of the thermodynamic algorithm (Algorithm \ref{alg:2}) for solving quadratic programs: Classification using support vector machines (SVMs), Porfolio optimization and simulation of nonlinear resisitve networks. For the first application (SVMs) we also provide numerical results which showcase the potential advantage of thermodynamic approach over state-of-the-art digital techniques.

\subsection{Support Vector Machines}
Support Vector Machines (SVMs) are versatile supervised learning algorithms used for classification and regression tasks \cite{cortes1995support}. They have a wide range of applications, notably in image recognition \cite{decoste2002training}, anomaly detection \cite{scholkopf1999support} or text classification \cite{joachims1998text}. The main idea behind SVMs is to find an optimal hyperplane that separates the data into different classes while maximizing the margin between the classes.

\begin{figure*}[t]
    \centering
    \includegraphics[width=0.9\linewidth]{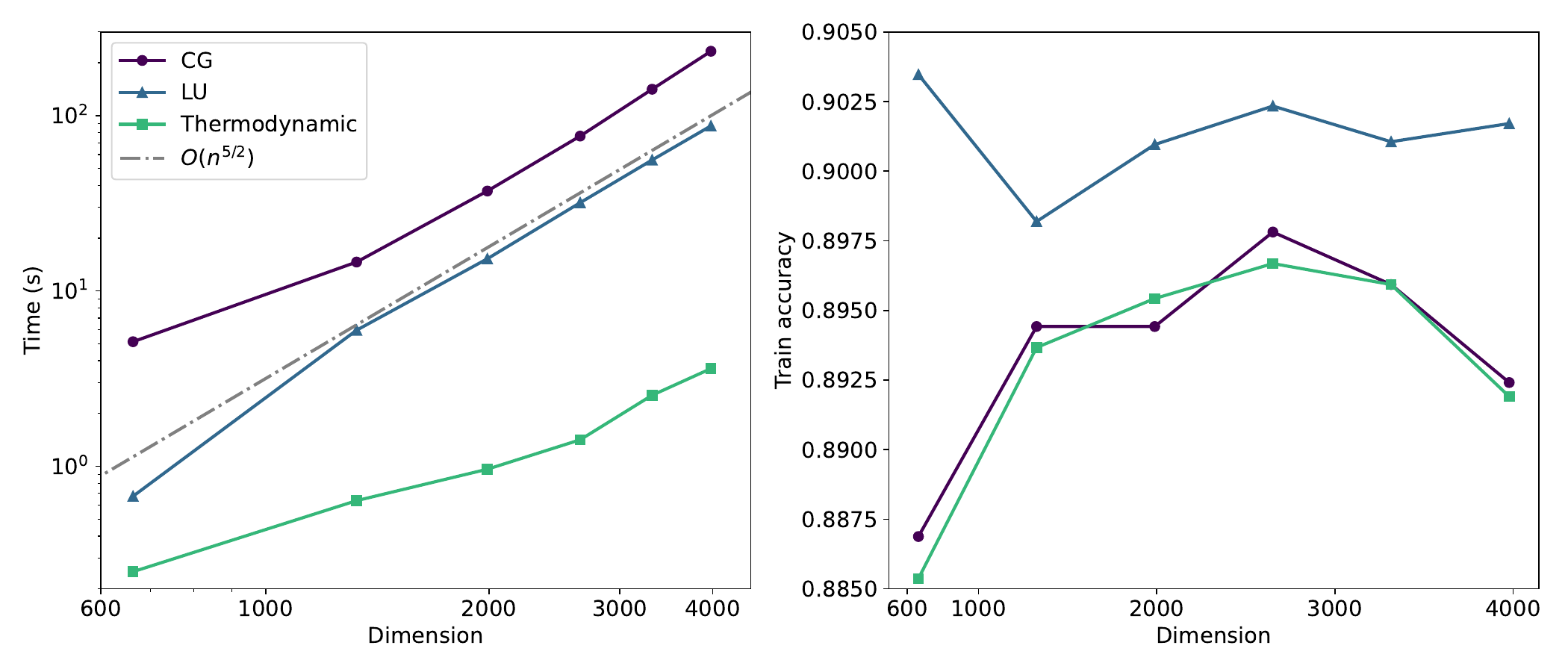}
    \caption{\textbf{Comparison of digital and thermodynamic approaches to training SVMs.} The left panel compares the training times when different subroutines for the IPM are used. The thermodynamic subroutine runtime is estimated with a timing model, detailed in Appendix~\ref{app:hardware}. The right panel shows the associated training accuracy of the model, showing a minimal degradation in solution quality. Here regularization is added for both the CG and thermodynamic IPMs, to stabilize the linear system $\tilde{J}_k \bm{\Delta r}_k = \bm{\tilde{v}}_k$, with parameter $\lambda = 0.1$. The simulations were performed on an Nvidia A100 GPU.}
    \label{fig:svm}
\end{figure*}

Consider a binary classification dataset $\mathcal{D} = \{\bm{x}_i, y_i\}_{i=1}^D$ where $\bm{x}_i \in \mathbb{R}^n$ and $y_i \in \{-1, 1\}$ for $i = 1, \ldots, D$. The separating hyperplane $\mathcal{W}$ is specified by a linear equation
\begin{align}
    \mathcal{W} := \{\bm{x} \,\, : \, \bm{w}^{\top} \bm{x} + b = 0 \},
\end{align}
where $\bm{w}$ is an vector of weights and $b$ is the bias term. The hyperplane $\mathcal{W}$ separates the dataset into two classes. The distance of a single point $\bm{x}_i$ from the hyperplane is given by $\gamma_i := y_i(\bm{w}^{\top} \bm{x}_i + b) / \norm{\bm{w}}$. The margin with respect to the entire dataset $\mathcal{D}$ is defined as $\gamma := \min_{i = 1, \ldots, N} \gamma_i$. The goal of the SVM algorithm is to find the weights $\bm{w}$ and the bias $b$ such that the resulting hyperplane maximizes the separation between the positive and the negative samples as specified by $\gamma$. This is achieved by introducing additional constraints of the form $y_i(\bm{w}^{\top} \bm{x} + b) \geq 1$. Data points which saturate the above inequalities are called \emph{support vectors}. These points achieve the minimal distance from the hyperplane $\mathcal{W}$ equal to the margin $\gamma$. 
The distance between support vectors and the hyperplane is hence given by $\gamma = 1 / \norm{\bm{w}}$.  Therefore maximizing the margin subject to the constraints is equivalent to solving the following problem 
\begin{align}
    \min_{{\bm{w}, b}}& \quad \norm{\bm{w}}  
    \\
    \text{subject to}& \quad    y_i(\bm{w}^{\top}\bm{x}_i + b) \geq 1 \quad \text{for all} \quad \{\bm{x}_i, y_i\} \in \mathcal{D}. 
\end{align}
The above problem of determining the separating hyperplane can be cast into a quadratic program in the standard form. For that we define a (regularized) Gram matrix $Q$ with elements defined as $Q_{ij}~=~y_i y_j \bm{x}_i^{\top} \bm{x}_j + \lambda \mathbb{I}$ and write $\bm{w} = \sum_{i} \alpha_i y_i \bm{x}_i$, where $\bm{\alpha}$ is a vector of Lagrange multipliers associated with the constraints. The rewritten problem reads
\begin{align}
    \min_{\bm{\alpha} \in \mathbb{R}^N}& \quad V(\bm{\alpha}) := \frac{1}{2} \bm{\alpha}^\top Q \bm{\alpha} + \bm{1}^\top \bm{\alpha}  \\
    \text{subject to}& \quad    \bm{y}^T \bm{\alpha} = 0, \ \nonumber\\
    &\quad \bm{\alpha} \geq 0. \nonumber
\end{align}
By comparing the above with Eq. \eqref{eq:qprog} we therefore see that this case corresponds to taking $\bm{c} = \bm{1}$, $A = \bm{y}^{\top}$ with $\bm{y} = [y_1, \ldots, y_N]^{\top}$ and $\bm{b} = 0$.

In Fig.~\ref{fig:svm}, we compare three different approaches to training an SVM model using an interior point algorithm on the UCI Breast cancer dataset~\cite{breast_cancer_wisconsin_17}. In order to observe scaling laws, we augment the data by duplicating points with added random Gaussian noise. Specifically, we train the classifier using digital linear solvers: LU decomposition, conjugate gradients (CG) 
as well as a thermodynamic algorithm~(Algorithm \ref{alg:2}), in which we simulate running the linear solver on specialized hardware. Based on realistic hardware assumptions detailed in Appendix~\ref{app:hardware}, we predict that the training time of the thermodynamic model significantly outperforms the digital one even for a moderate size of the dataset, reaching a speedup of $10-30\times $ for dimensions in the thousands. Moreover, we observe that for moderate dimensions the runtime scaling with respect to the dimension $n$ of the CG algorithm is close to the asymptotic value $\mathcal{O}(n^{5/2})$. The LU approach seems to offer a similar time complexity, despite a slightly worse theoretical scaling (see Tab. \ref{tab:complexities}), likely due to high parallelization on the GPU. In the same time, the thermodynamic algorithm outperforms other algorithms in terms of both actual runtime and theoretical complexity. The training accuracy of CG and the thermodynamic algorithm are slighly worse than LU decomposition due to the need to add regularization to the linear system involving $\tilde{J}_k$ (which has worse conditioning than $J_k$). However, they remain within $\sim1-2\%$ error of the LU algorithm training error. The runtime for the thermodynamic algorithm is estimated using realistic equilibration timescales for electronic systems and includes input/output times (including digital-to-analog conversion), see Appendix \ref{app:hardware} for a detailed discussion. The simulations were performed with the $\texttt{thermox}$ package~\cite{duffield2024thermox}.

\subsection{Portfolio optimization}

Portfolio optimization is a mathematical technique used to determine the optimal allocation of assets in an investment portfolio. It aims to maximize portfolio returns while minimizing risks, subject to certain constraints.

Let $\bm{x}$ be a vector of size $N$, representing the proportion of wealth invested in each of the $N$ available assets (e.g., $x_i$ represents the proportion invested in asset $i$). Each asset has a known rate of return of $r_i$ so that the expected return is given by $\bm{x}^{\top}\bm{r}$ with $\bm{r} = (r_1, \ldots, r_N)$. The portfolio variance is a quadratic function of $\bm{x}$ given by $V(\bm{x}) = \frac{1}{2}\bm{x}^{\top} Q \bm{x}$, where $Q \in \mathbb{R}^{N\times N}$ is the covariance matrix of asset returns given by $Q = (Q_{ij})$ with $Q_{ij} := \text{var}(r_i, r_j)$.

Portfolio optimization seeks to minimize the risk (variance) associated with different allocations of wealth among available assets. The objective is to find the optimal allocation that achieves a target expected return $R$, while minimizing the portfolio's overall risk. This optimization problem can be formulated as a quadratic programming problem:

\begin{align}
    \min_{\bm{x} \in \mathbb{R}^N}& \quad V(\bm{x}) := \frac{1}{2} \bm{x}^\top Q \bm{x}  \\
    \text{subject to}& \quad    \bm{x}^T\bm{r} \geq R, \ \nonumber\\
    & \quad     \bm{1}^T \bm{x} = 1, \nonumber \\
    &\quad \bm{x} \geq 0. \nonumber
\end{align}

The problem of portfolio optimization can be therefore efficiently solved using thermodynamic hardware, as discussed in Section \ref{sec:thermo_quad}.

\subsection{Simulation of Nonlinear Resistive Networks}

Nonlinear resistor networks (NRNs) are emerging as a powerful alternative to traditional GPU-based neural networks \cite{scellier2023universal}. Their key advantage lies in directly implementing neural network dynamics in an analog fashion. This analog approach translates to significantly lower power consumption compared to digital implementations. Additionally, NRNs can be trained efficiently using local techniques such as equilibrium propagation \cite{scellier2017equilibrium}.

However, efficiently simulating NRNs presents a major challenge, hindering our ability to assess their scalability. Recently, it has been shown that simulating NRNs can be formulated as a quadratic program \cite{scellier2024fast}. Here we briefly review this promising approach and show how a thermodynamic algorithm can be used to solve such programs for efficient simulation of resistive networks.

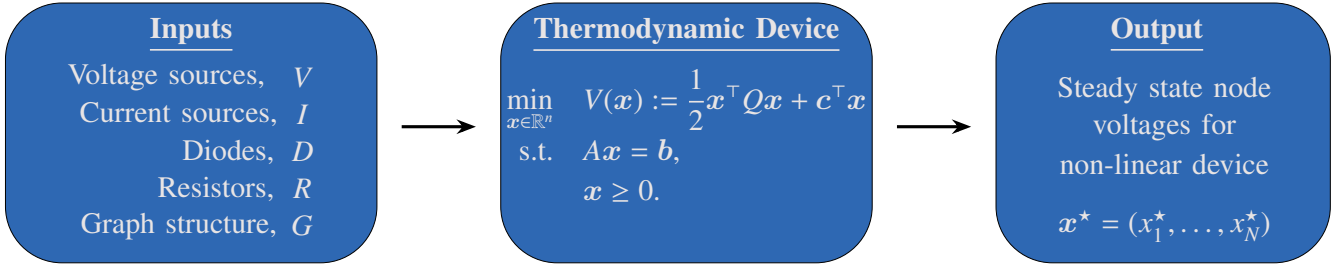
\begin{figure*}[t]
\centering
\resizebox{1\textwidth}{!}{%
\begin{circuitikz}
\tikzstyle{every node}=[font=\LARGE]
\draw [ fill={rgb,255:red,46; green,104; blue,179} , line width=0.2pt , rounded corners = 32.4] (0,10.75) rectangle (7.5,5.5);
\draw [ fill={rgb,255:red,46; green,104; blue,179} , line width=0.2pt , rounded corners = 32.4] (10,10.75) rectangle (17.5,5.5);
\draw [ fill={rgb,255:red,46; green,104; blue,179} , line width=0.2pt , rounded corners = 32.4] (20,10.75) rectangle (26.75,5.5);
\draw [line width=1.8pt, ->, >=Stealth] (8,8.25) -- (9.5,8.25);
\draw [line width=1.8pt, ->, >=Stealth] (18,8.25) -- (19.5,8.25);
\node [font=\LARGE, color={rgb,255:red,232; green,228; blue,228}] at (13.75,10.1) {\textbf{\underline{Thermodynamic Device}}};
\node [font=\LARGE, color={rgb,255:red,232; green,228; blue,228}] at (13.75,8.1) {
\begin{minipage}{0.7\linewidth}
  \begin{align} \nonumber
       \min_{\bm{x} \in \mathbb{R}^n}& \quad V(\bm{x}) := \frac{1}{2} \bm{x}^\top Q \bm{x} + \bm{c}^\top \bm{x}  \\
    \text{s.t.}& \quad    A \bm{x} = \bm{b}, \ \nonumber\\
    &\quad \bm{x} \geq 0. \nonumber
\end{align}
\end{minipage}
};;
\node [font=\LARGE, color={rgb,255:red,232; green,228; blue,228}] at (3.75,10.1) {\textbf{\underline{Inputs}}};
\node [font=\LARGE, color={rgb,255:red,232; green,228; blue,228}] at (23.25,10.1) {\textbf{\underline{Output}}};
\node [font=\LARGE, color={rgb,255:red,232; green,228; blue,228}] at (3.25,9.25) {Voltage sources,};
\node [font=\LARGE, color={rgb,255:red,232; green,228; blue,228}] at (6,9.25) {$V$};
\node [font=\LARGE, color={rgb,255:red,232; green,228; blue,228}] at (3.5,8.5) {Current sources, };
\node [font=\LARGE, color={rgb,255:red,232; green,228; blue,228}] at (6,8.5) {$I$};
\node [font=\LARGE, color={rgb,255:red,232; green,228; blue,228}] at (4.25,7) {Resistors, };
\node [font=\LARGE, color={rgb,255:red,232; green,228; blue,228}] at (6,7.75) {$D$};
\node [font=\LARGE, color={rgb,255:red,232; green,228; blue,228}] at (6,7) {$R$};
\node [font=\LARGE, color={rgb,255:red,232; green,228; blue,228}] at (3.5,6.25) {Graph structure,};
\node [font=\LARGE, color={rgb,255:red,232; green,228; blue,228}] at (4.5,7.75) {Diodes, };
\node [font=\LARGE, color={rgb,255:red,232; green,228; blue,228}] at (23.4,7.5) {non-linear device};
\node [font=\LARGE, color={rgb,255:red,232; green,228; blue,228}] at (23.4,8.25) {voltages for};
\node [font=\LARGE, color={rgb,255:red,232; green,228; blue,228}] at (23.4,9) {Steady state node};
\node [font=\LARGE, color={rgb,255:red,232; green,228; blue,228}] at (23.4,6.25) {$\bm{x}^{\star} = (x_1^{\star}, \ldots, x_N^{\star})$};
\node [font=\LARGE, color={rgb,255:red,232; green,228; blue,228}] at (6,6.25) {$G$};
\end{circuitikz}
}%
\caption{\justifying \textbf{Quadratic programming approach to resisitve networks.} The parameters of the resistive networks (i.e. voltage sources $V$, current sources $I$, diodes $D$, resistors $R$ and the graph structure $G$ of the network) are encoded in the parameters of a quadratic program ($A$, $\bm{b}$ and $\bm{c}$). The thermodynamic interior point algorithm (Algorithm \ref{alg:2}) is then used to solve the optimization problem and produces a close-to-optimal solution $\bm{x}^{\star}$. The solution encodes the information about the steady state  configuration (node voltages) of the resistive network.}
\label{fig:res_nets}
\end{figure*}

Let us consider an (ideal) nonlinear resistive network with $d$ nodes with $\bm{x} = (x_1, \ldots, x_d)$ denoting the node electrical potentials. A nonlinear resistive network can be represented as a graph in which every branch contains a unique (ideal) element: a voltage source (V), a current source (C), a diode (D) or a linear resistor (R) (see Ref. \cite{scellier2023universal} for further details). The branches of the graph are denoted with $\mathcal{B} = \mathcal{B}_{V} \cup \mathcal{B}_{C} \cup \mathcal{B}_{D} \cup \mathcal{B}_{R}$, where $\mathcal{B}_{r}$ is a set of branches containing only elements of type $r$. For every branch $(j,k) \in \mathcal{B}_{vs}$ one denotes with $I_{jk}$ the current through the current source between nodes $j$ and $k$. Similarly $V_{ij}$ denotes the voltage and $G_{jk}$ the conductance  between these nodes. Finally, the last element, the diode, can be in one of two states: the off-state where it behaves like an open switch (no current can flow through it), or the on-state (current can flow freely through it).

A configuration of branch voltages and branch currents that satisfies Kirchhoff’s current and voltage laws is known as the \emph{steady state configuration}. Such a configuration minimizes the potential function $V(\bm{x})$ of the network. The steady state $\bm{x}^{\star}$ can be found by solving a quadratic program corresponding to such a network \cite{scellier2024fast}, i.e.

\begin{align} \label{eq:nrn_quad}
       \min_{\bm{x} \in \mathbb{R}^n}& \quad V(\bm{x}) := \frac{1}{2} \bm{x}^\top Q \bm{x} + \bm{c}^\top \bm{x}  \\
    \text{subject to}& \quad    A \bm{x} = \bm{b}, \ \nonumber\\
    &\quad \bm{x} \geq 0. \nonumber
\end{align}
In the current notation, the parameters of the quadratic program specifying the steady state configuration are given by
\begin{align}
    Q_{ij} &= \begin{cases}
        g_{jk + kj} \quad &\text{if} \quad j = k \quad \text{and} \quad (j, k) \in \mathcal{B}_{R}, \\
        -g_{jk} \quad &\text{if} \quad j \neq k \quad \text{and} \quad (j, k) \in \mathcal{B}_{R}, \\
        \,\,\, 0 &\text{else},
    \end{cases}\\
    c_j &= \begin{cases}
        I_{jk} - I_{kj} \quad &\text{if} \quad (j, k) \in \mathcal{B}_{R}, \\
        0 &\text{else}.
    \end{cases}
\end{align}
Solving the quadratic program from Eq. \eqref{eq:nrn_quad} using the thermodynamic algorithm (Algorithm \ref{alg:2}) therefore provides an alternative method of computing the steady state $\bm{x}^{\star}$ of an ideal nonlinear resistive network (see Fig. \ref{fig:res_nets}).

\section*{Discussion}
We have introduced a new method for solving quadratic programming problems using thermodynamic computing hardware. Our approach leverages the natural relaxation properties of thermodynamic systems to efficiently find optimal solutions. 

Specifically, we adapted the interior point method by replacing its most computationally intensive step—-solving the system of non-linear equations defined by the KKT 
optimality conditions—-with a thermodynamic subroutine that solves the linearized KKT system. We also explored three applications of this hybrid digital-thermodynamic algorithm: portfolio optimization, machine learning (support vector machines), and simulating non-linear resistive networks. Focusing on support vector machine classification, we demonstrated a potential thermodynamic advantage in training time for moderately-sized datasets. Note that our method may be employed with other hardware accelerators, such as those put forward in~\cite{kalinin2023analog} that rely on optical components.

This work opens up several interesting avenues for further exploration. One such direction is to extend our analysis to more complex convex optimization problems that can be addressed using interior point methods. Semidefinite programs \cite{boyd2004convex}, widely used in quantum information theory, are a prime example. Whether a thermodynamic solver would offer an advantage in these cases requires further analysis: One would need to analyse the specific structure of the KKT conditions in this case and check whether the digital step of computing the symmetrized Jacobian of the problem could be simplified similarly as in the case of quadratic programming.

Another interesting direction would be to apply Algorithm \ref{alg:2} to solve general optimization problems. We emphasize that any sufficiently regular nonlinear optimization problem can usually be effectively approximated using the technique of Sequential Quadratic Programming (SQP). SQP methods iteratively approximate the original nonlinear problem with a sequence of quadratic subproblems, similar to those in equation \eqref{eq:qprog}, and solve these subproblems until convergence.  It would be interesting to quantify the runtime performance of thermodynamic implementations of such SQP algorithms, and compare it to digital algorithms running on CPUs or GPUs.

A further promising direction would be to explore connections between thermodynamic computing and thermodynamically inspired optimization problems. Many heuristics used widely in both academia and industry rely on algorithms based on thermodynamic principles, such as simulated annealing \cite{van1987simulated} or the matrix multiplicative weights method \cite{Arora2016}.  These algorithms often involve steps that resemble physical thermalization, which is also computationally expensive when implemented digitally. It would be interesting to investigate whether these thermodynamically inspired algorithms could be implemented using actual thermodynamic processes, potentially offering advantages in terms of time or space requirements compared to digital implementations.




\bibliography{references,thermo}

\begin{thebibliography}{10}
\providecommand{\url}[1]{#1}
\csname url@samestyle\endcsname
\providecommand{\newblock}{\relax}
\providecommand{\bibinfo}[2]{#2}
\providecommand{\BIBentrySTDinterwordspacing}{\spaceskip=0pt\relax}
\providecommand{\BIBentryALTinterwordstretchfactor}{4}
\providecommand{\BIBentryALTinterwordspacing}{\spaceskip=\fontdimen2\font plus
\BIBentryALTinterwordstretchfactor\fontdimen3\font minus \fontdimen4\font\relax}
\providecommand{\BIBforeignlanguage}[2]{{%
\expandafter\ifx\csname l@#1\endcsname\relax
\typeout{** WARNING: IEEEtran.bst: No hyphenation pattern has been}%
\typeout{** loaded for the language `#1'. Using the pattern for}%
\typeout{** the default language instead.}%
\else
\language=\csname l@#1\endcsname
\fi
#2}}
\providecommand{\BIBdecl}{\relax}
\BIBdecl

\bibitem{Mohseni2022}
\BIBentryALTinterwordspacing
N.~Mohseni, P.~L. McMahon, and T.~Byrnes, ``Ising machines as hardware solvers of combinatorial optimization problems,'' \emph{Nature Reviews Physics}, vol.~4, no.~6, p. 363–379, May 2022. [Online]. Available: \url{http://dx.doi.org/10.1038/s42254-022-00440-8}
\BIBentrySTDinterwordspacing

\bibitem{Bybee2023}
\BIBentryALTinterwordspacing
C.~Bybee, D.~Kleyko, D.~E. Nikonov, A.~Khosrowshahi, B.~A. Olshausen, and F.~T. Sommer, ``Efficient optimization with higher-order ising machines,'' \emph{Nature Communications}, vol.~14, no.~1, Sep. 2023. [Online]. Available: \url{http://dx.doi.org/10.1038/s41467-023-41214-9}
\BIBentrySTDinterwordspacing

\bibitem{Tanahashi2019}
\BIBentryALTinterwordspacing
K.~Tanahashi, S.~Takayanagi, T.~Motohashi, and S.~Tanaka, ``Application of ising machines and a software development for ising machines,'' \emph{Journal of the Physical Society of Japan}, vol.~88, no.~6, p. 061010, Jun. 2019. [Online]. Available: \url{http://dx.doi.org/10.7566/JPSJ.88.061010}
\BIBentrySTDinterwordspacing

\bibitem{chou2019analog}
J.~Chou, S.~Bramhavar, S.~Ghosh, and W.~Herzog, ``Analog coupled oscillator based weighted ising machine,'' \emph{Scientific reports}, vol.~9, no.~1, p. 14786, 2019.

\bibitem{yamamoto2020coherent}
Y.~Yamamoto, T.~Leleu, S.~Ganguli, and H.~Mabuchi, ``Coherent ising machines—quantum optics and neural network perspectives,'' \emph{Applied Physics Letters}, vol. 117, no.~16, p. 160501, 2020.

\bibitem{wang2019oim}
T.~Wang and J.~Roychowdhury, ``Oim: Oscillator-based ising machines for solving combinatorial optimisation problems,'' in \emph{Unconventional Computation and Natural Computation: 18th International Conference, UCNC 2019, Tokyo, Japan, June 3--7, 2019, Proceedings 18}.\hskip 1em plus 0.5em minus 0.4em\relax Springer, 2019, pp. 232--256.

\bibitem{patel2024pass}
S.~Patel, P.~Canoza, A.~Datar, S.~Lu, C.~Garg, and S.~Salahuddin, ``Pass: An asynchronous probabilistic processor for next generation intelligence,'' \emph{arXiv preprint arXiv:2409.10325}, 2024.

\bibitem{Hauke_2020}
\BIBentryALTinterwordspacing
P.~Hauke, H.~G. Katzgraber, W.~Lechner, H.~Nishimori, and W.~D. Oliver, ``Perspectives of quantum annealing: methods and implementations,'' \emph{Reports on Progress in Physics}, vol.~83, no.~5, p. 054401, May 2020. [Online]. Available: \url{http://dx.doi.org/10.1088/1361-6633/ab85b8}
\BIBentrySTDinterwordspacing

\bibitem{Camsari_2019}
K.~Y. Camsari, B.~M. Sutton, and S.~Datta, ``p-bits for probabilistic spin logic,'' \emph{Appl. Phys. Rev.}, vol.~6, no.~1, p. 011305, 2019.

\bibitem{kaiser2022life}
J.~Kaiser, S.~Datta, and B.~Behin-Aein, ``Life is probabilistic—why should all our computers be deterministic? computing with p-bits: Ising solvers and beyond,'' in \emph{2022 International Electron Devices Meeting (IEDM)}.\hskip 1em plus 0.5em minus 0.4em\relax IEEE, 2022, pp. 21--4.

\bibitem{chowdhury2023full}
S.~Chowdhury, A.~Grimaldi, N.~A. Aadit, S.~Niazi, M.~Mohseni, S.~Kanai, H.~Ohno, S.~Fukami, L.~Theogarajan, G.~Finocchio \emph{et~al.}, ``A full-stack view of probabilistic computing with p-bits: devices, architectures, and algorithms,'' \emph{IEEE Journal on Exploratory Solid-State Computational Devices and Circuits}, vol.~9, no.~1, pp. 1--11, 2023.

\bibitem{aadit2023all}
N.~A. Aadit, S.~Nikhar, S.~Kannan, S.~Chowdhury, and K.~Y. Camsari, ``All-to-all reconfigurability with sparse ising machines: the xorsat challenge with p-bits,'' \emph{arXiv preprint arXiv:2312.08748}, 2023.

\bibitem{kalinin2023analogiterativemachineaim}
\BIBentryALTinterwordspacing
K.~P. Kalinin, G.~Mourgias-Alexandris, H.~Ballani, N.~G. Berloff, J.~H. Clegg, D.~Cletheroe, C.~Gkantsidis, I.~Haller, V.~Lyutsarev, F.~Parmigiani, L.~Pickup, and A.~Rowstron, ``Analog iterative machine (aim): using light to solve quadratic optimization problems with mixed variables,'' 2023. [Online]. Available: \url{https://arxiv.org/abs/2304.12594}
\BIBentrySTDinterwordspacing

\bibitem{conte2019thermodynamic}
T.~Conte, E.~DeBenedictis, N.~Ganesh, T.~Hylton, J.~P. Strachan, R.~S. Williams, A.~Alemi, L.~Altenberg, G.~Crooks, J.~Crutchfield \emph{et~al.}, ``Thermodynamic computing,'' \emph{arXiv preprint arXiv:1911.01968}, 2019.

\bibitem{wiredarticle}
W.~Knight, ``{C}hat{G}{P}{T}'s {H}unger for {E}nergy {C}ould {T}rigger a {G}{P}{U} {R}evolution --- wired.com,'' \url{https://www.wired.com/story/fast-forward-chatgpt-hunger-energy-gpu-revolution}, 2024.

\bibitem{hylton2020thermodynamic}
T.~Hylton, ``Thermodynamic neural network,'' \emph{Entropy}, vol.~22, no.~3, p. 256, 2020.

\bibitem{lipka2024thermodynamic}
P.~Lipka-Bartosik, M.~Perarnau-Llobet, and N.~Brunner, ``Thermodynamic computing via autonomous quantum thermal machines,'' \emph{Science Advances}, vol.~10, no.~36, p. eadm8792, 2024.

\bibitem{hylton2022thermodynamic}
T.~Hylton, ``Thermodynamic state machine network,'' \emph{Entropy}, vol.~24, no.~6, p. 744, 2022.

\bibitem{donatella2024thermodynamic}
K.~Donatella, S.~Duffield, M.~Aifer, D.~Melanson, G.~Crooks, and P.~J. Coles, ``Thermodynamic natural gradient descent,'' \emph{arXiv preprint arXiv:2405.13817}, 2024.

\bibitem{aifer2024npj}
M.~Aifer, K.~Donatella, M.~H. Gordon, S.~Duffield, T.~Ahle, D.~Simpson, G.~Crooks, and P.~J. Coles, ``Thermodynamic linear algebra,'' \emph{npj Unconventional Computing}, vol.~1, no.~1, p.~13, 2024.

\bibitem{duffield2023thermodynamic}
S.~Duffield, M.~Aifer, G.~Crooks, T.~Ahle, and P.~J. Coles, ``Thermodynamic matrix exponentials and thermodynamic parallelism,'' \emph{arXiv preprint arXiv:2311.12759}, 2023.

\bibitem{coles2023thermodynamic}
P.~J. Coles, C.~Szczepanski, D.~Melanson, K.~Donatella, A.~J. Martinez, and F.~Sbahi, ``Thermodynamic {AI} and the fluctuation frontier,'' in \emph{2023 IEEE International Conference on Rebooting Computing (ICRC)}.\hskip 1em plus 0.5em minus 0.4em\relax IEEE, 2023, pp. 1--10.

\bibitem{aifer2024thermodynamic}
M.~Aifer, S.~Duffield, K.~Donatella, D.~Melanson, P.~Klett, Z.~Belateche, G.~Crooks, A.~J. Martinez, and P.~J. Coles, ``Thermodynamic bayesian inference,'' \emph{arXiv preprint arXiv:2410.01793}, 2024.

\bibitem{melanson2023thermodynamic}
D.~Melanson, M.~A. Khater, M.~Aifer, K.~Donatella, M.~H. Gordon, T.~Ahle, G.~Crooks, A.~J. Martinez, F.~Sbahi, and P.~J. Coles, ``Thermodynamic computing system for {AI} applications,'' \emph{arXiv preprint arXiv:2312.04836}, 2023.

\bibitem{aifer2024error}
M.~Aifer, D.~Melanson, K.~Donatella, G.~Crooks, T.~Ahle, and P.~J. Coles, ``Error mitigation for thermodynamic computing,'' \emph{arXiv preprint arXiv:2401.16231}, 2024.

\bibitem{sun2019solving}
Z.~Sun, G.~Pedretti, E.~Ambrosi, A.~Bricalli, W.~Wang, and D.~Ielmini, ``Solving matrix equations in one step with cross-point resistive arrays,'' \emph{Proceedings of the National Academy of Sciences}, vol. 116, no.~10, pp. 4123--4128, 2019.

\bibitem{vanderbei1998linear}
R.~J. Vanderbei, ``Linear programming: foundations and extensions,'' \emph{Journal of the Operational Research Society}, vol.~49, no.~1, pp. 94--94, 1998.

\bibitem{SINUANYSTERN20231069}
\BIBentryALTinterwordspacing
Z.~Sinuany-Stern, ``Foundations of operations research: From linear programming to data envelopment analysis,'' \emph{European Journal of Operational Research}, vol. 306, no.~3, pp. 1069--1080, 2023. [Online]. Available: \url{https://www.sciencedirect.com/science/article/pii/S0377221722008578}
\BIBentrySTDinterwordspacing

\bibitem{GONDZIO2012587}
\BIBentryALTinterwordspacing
J.~Gondzio, ``Interior point methods 25 years later,'' \emph{European Journal of Operational Research}, vol. 218, no.~3, pp. 587--601, 2012. [Online]. Available: \url{https://www.sciencedirect.com/science/article/pii/S0377221711008204}
\BIBentrySTDinterwordspacing

\bibitem{boyd2004convex}
S.~P. Boyd and L.~Vandenberghe, \emph{Convex optimization}.\hskip 1em plus 0.5em minus 0.4em\relax Cambridge university press, 2004.

\bibitem{hamming2012numerical}
R.~Hamming, \emph{Numerical methods for scientists and engineers}.\hskip 1em plus 0.5em minus 0.4em\relax Courier Corporation, 2012.

\bibitem{boltzmann1871einige}
L.~E. Boltzmann, \emph{Einige allgemeine S{\"a}tze {\"u}ber W{\"a}rmegleichgewicht}.\hskip 1em plus 0.5em minus 0.4em\relax K. Akad. der Wissensch., 1871.

\bibitem{gondzio2013convergence}
J.~Gondzio, ``Convergence analysis of an inexact feasible interior point method for convex quadratic programming,'' \emph{SIAM Journal on Optimization}, vol.~23, no.~3, pp. 1510--1527, 2013.

\bibitem{mizuno1999global}
S.~Mizuno and F.~Jarre, ``Global and polynomial-time convergence of an infeasible-interior-point algorithm using inexact computation.'' \emph{Mathematical Programming}, vol.~84, no.~1, 1999.

\bibitem{cortes1995support}
C.~Cortes and V.~Vapnik, ``Support-vector networks,'' \emph{Machine learning}, vol.~20, pp. 273--297, 1995.

\bibitem{decoste2002training}
D.~DeCoste and B.~Sch{\"o}lkopf, ``Training invariant support vector machines,'' \emph{Machine learning}, vol.~46, pp. 161--190, 2002.

\bibitem{scholkopf1999support}
B.~Sch{\"o}lkopf, R.~C. Williamson, A.~Smola, J.~Shawe-Taylor, and J.~Platt, ``Support vector method for novelty detection,'' \emph{Advances in neural information processing systems}, vol.~12, 1999.

\bibitem{joachims1998text}
T.~Joachims, ``Text categorization with support vector machines: Learning with many relevant features,'' in \emph{European conference on machine learning}.\hskip 1em plus 0.5em minus 0.4em\relax Springer, 1998, pp. 137--142.

\bibitem{breast_cancer_wisconsin_17}
\BIBentryALTinterwordspacing
O.~M. William~Wolberg, ``Breast cancer wisconsin (diagnostic),'' 1993. [Online]. Available: \url{https://archive.ics.uci.edu/dataset/17}
\BIBentrySTDinterwordspacing

\bibitem{duffield2024thermox}
S.~Duffield, K.~Donatella, and D.~Melanson, ``thermox: Exact ou processes with {JAX},'' \url{https://github.com/normal-computing/thermox}, 2024.

\bibitem{scellier2023universal}
B.~Scellier and S.~Mishra, ``A universal approximation theorem for nonlinear resistive networks,'' \emph{arXiv preprint arXiv:2312.15063}, 2023.

\bibitem{scellier2017equilibrium}
B.~Scellier and Y.~Bengio, ``Equilibrium propagation: Bridging the gap between energy-based models and backpropagation,'' \emph{Frontiers in computational neuroscience}, vol.~11, p.~24, 2017.

\bibitem{scellier2024fast}
B.~Scellier, ``A fast algorithm to simulate nonlinear resistive networks,'' \emph{arXiv preprint arXiv:2402.11674}, 2024.

\bibitem{kalinin2023analog}
K.~P. Kalinin, G.~Mourgias-Alexandris, H.~Ballani, N.~G. Berloff, J.~H. Clegg, D.~Cletheroe, C.~Gkantsidis, I.~Haller, V.~Lyutsarev, F.~Parmigiani \emph{et~al.}, ``Analog iterative machine (aim): using light to solve quadratic optimization problems with mixed variables,'' \emph{arXiv preprint arXiv:2304.12594}, 2023.

\bibitem{van1987simulated}
P.~J. Van~Laarhoven, E.~H. Aarts, P.~J. van Laarhoven, and E.~H. Aarts, \emph{Simulated annealing}.\hskip 1em plus 0.5em minus 0.4em\relax Springer, 1987.

\bibitem{Arora2016}
\BIBentryALTinterwordspacing
S.~Arora and S.~Kale, ``A combinatorial, primal-dual approach to semidefinite programs,'' \emph{J. ACM}, vol.~63, no.~2, may 2016. [Online]. Available: \url{https://doi.org/10.1145/2837020}
\BIBentrySTDinterwordspacing

\end{thebibliography}

\newpage
\onecolumn
\section*{Supplementary Materials}
\setcounter{section}{0} 
\section{Interior Point Methods}
\label{sec:IPMs}
The starting point for any IPM algorithm is to transform a given optimization problem into a set of (potentially non-linear) equations. This is achieved by first replacing the non-negativity constraints $\bm{x} \geq 0$ with the an appropriate penalty function and forming the associated Lagrangian. Differentiating this Lagrangian with respect to the problem variables gives the first-order optimality conditions known as Karush-Kuhn-Tucker (KKT) conditions. These conditions are both sufficient and necessary for a global solution of Eq. \eqref{eq:qprog}.

For the problem \eqref{eq:qprog} one can write the KKT conditions as  \cite{boyd2004convex}:
\begin{subequations}
\label{eq:kkt}
\begin{align}
    \label{eq:kkt1}
    Q \bm{x} - A^{\top} \bm{y} - \bm{z} + \bm{c} &= 0, \\
    \label{eq:kkt2}
    A \bm{x} - \bm{b} &= 0, \\
    \label{eq:kkt3}
    x_i z_i &= \mu \quad \text{for} \quad i = 1, \ldots, n, \\
    \label{eq:kkt4}
    \bm{x}, \bm{z} &\geq 0. 
\end{align}
\end{subequations}
In the above we introduced slack variables $\bm{y}\in \mathbb{R}^m$ and $\bm{z} \in \mathbb{R}^n$. In IPM the standard complementarity slackness condition \eqref{eq:kkt3} is replaced by its perturbed version with a barrier term $\mu > 0$. The idea of IPM is to drive the barrier term $\mu$ to zero and gradually reveal the activity of the problem variables. In other words, as $\mu \rightarrow 0$ more importance is given to optimality over feasibility. This is achieved by applying the Newton's method to the system of nonlinear equations \eqref{eq:kkt}. That is, instead of solving this system exactly, the method starts from an iterate in the interior of
the feasible region $\bm{r}_k := (\bm{x}_k, \bm{y}_k, \bm{z}_k)$ and 
generates iterates given by
\begin{align}   \label{eq:central_path_app}
    \bm{r}_{k+1} = \bm{r}_k + \alpha (\Delta \bm{x}_k, \Delta \bm{y}_k, \Delta \bm{z}_k),
\end{align}
where the scalar $\alpha$ is selected
such that $\bm{x}_{k+1} > 0$ and $\bm{z}_{k+1} > 0$ and $(\Delta \bm{x}_k, \Delta \bm{y}_k, \Delta \bm{z}_k)$ can be found by solving the linear system of equations
\begin{align}
    \label{eq:linearized}
    J_k \begin{bmatrix}
        \Delta \bm{x}_k \\
        \Delta \bm{y}_k \\
        \Delta \bm{z}_k
    \end{bmatrix} = \bm{v}_k, 
\end{align}
where $J_k$ is the Jacobian of the original system of equations \eqref{eq:kkt} and $\bm{v}_k \in \mathbb{R}^{2n + m}$. These quantities are given by  
\begin{align}
   J(\bm{x}_k, \bm{y}_k, \bm{z}_k) &= \begin{bmatrix}
    \label{eq:ipm_lin}
        -Q & A^{\top} & I \\
        A & 0 &0 \\
        Z_k &0 &X_k
    \end{bmatrix}, \\ \label{eq:ipm_vec} \bm{v}_k &=
    \begin{bmatrix}
        Q \bm{x}_k - A^{\top} \bm{y}_k - \bm{z}_k + \bm{c} \\
        -A\bm{x}_k + \bm{b} \\
        -X_k Z_k \bm{e} + \mu_k \sigma_k \bm{e},
    \end{bmatrix}
\end{align}
where $I$ is the identity matrix, $\sigma_{k} \in (0, 1)$ is a barrier parameter and $\bm{e}$ is the identity vector $(1, \ldots, 1)$. Moreover, for clarity of notation we introduced diagonal matrices $X_k = \text{diag}(x_1, \ldots, x_n)$ 

By iteratively solving Eq. \eqref{eq:linearized} one obtains the \emph{central path} $\{\bm{r}_k \in \mathbb{R}^{2n+m}\,|\, k = 0, 1, \ldots, N\}$ which converges to the solution of the KKT conditions \eqref{eq:kkt}. From this one can then easily deduce the optimal solution $\bm{x}^{\star}$ of the original quadratic program from Eq. \eqref{eq:qprog}. 

\section{Hardware implementation}
\label{app:hardware}
In this section we provide more details about the model we use to provide results about timing and accuracy on the SVM task. The thermodynamic solver may be implemented on various physical platforms, but the most natural choice for very large-scale integration is a network of passive electronic elements.

\begin{figure}
    \centering
    \includegraphics[width=0.7\textwidth]{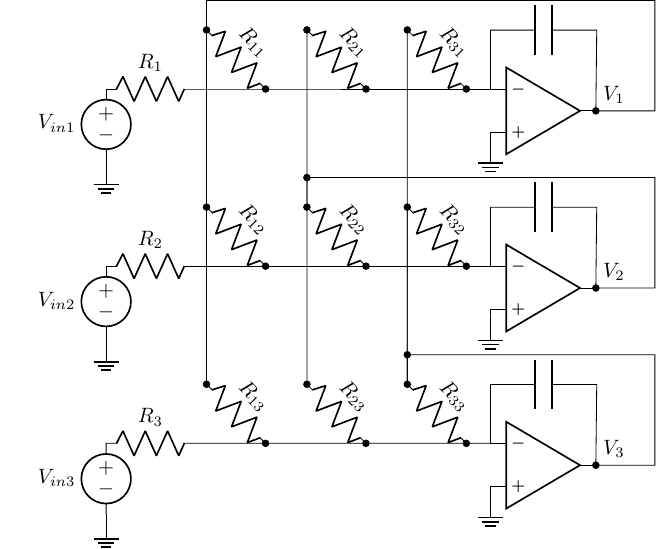}
    \caption{\textbf{Circuit diagram for the thermodynamic device in the case of a single resistor array implementation for a 3-dimensional problem.} The device is comprised of three voltage sources, each of which is connected to three resistors $\{R_i\}$ . Each of these resistors is connected to a line that goes into the negative pin of a different operational amplifier. A capacitor connects the negative pin of the operational amplifier to the operational amplifier's output port, where the voltage is denoted by $\{V_i\}$. The output ports of the operational amplifiers are connected to an array of $N\times N$ resistors $\{R_{ij}\}$ (nine here), each of which connects to the line going from the resistors $\{R_i\}$ to the negative pins of the operational amplifiers. This feedback loop enables the circuit to run a differential equation whose steady-state corresponds to the solution of a linear system of equations.}
    \label{fig:resistor_array}
\end{figure}

We consider a single resistor array described by the schematic shown in Fig.~\ref{fig:resistor_array}. (See also Refs.~\cite{donatella2024thermodynamic,sun2019solving} for discussion of similar resistor arrays.) By Kirchhoff’s current law, we obtain the equation of motion for the vector of voltages $V = (V_1, V_2, V_3)$ as:

\begin{equation}
    C\dot{V} = -\mathcal{G}V + R^{-1}V_{\mathrm{in}} + I_\mathrm{n},
\end{equation}

with $V_\mathrm{in} = (V_{\mathrm{in}1}, V_{\mathrm{in}2}, V_{\mathrm{in}3})$, $R = \mathrm{diag}(R_1, R_2, R_3)$, $C = \mathrm{diag}(C_1, C_2, C_3)$, and $I_\mathrm{n} = (I_{\mathrm{n}_1}, I_{\mathrm{n}_2}, I_{\mathrm{n}_3})$, which is the current noise vector and is assumed to be Gaussian with zero mean and variance $2\kappa_0$. In this case we have

\begin{equation}
    \mathcal{G} = \begin{pmatrix}\frac{1}{R_{11}} & \frac{1}{R_{21}} & \frac{1}{R_{31}}\\\frac{1}{R_{12}} & \frac{1}{R_{22}} & \frac{1}{R_{32}}\\\frac{1}{R_{13}} & \frac{1}{R_{23}} & \frac{1}{R_{33}}\end{pmatrix}. 
\end{equation} These resistor values need to be tunable such that different Jacobians and Hessians may be given as inputs throughout the training process (and for the hardware to be used for different problems). 

At steady state ($\dot{V} = 0$), the average voltage vector is $\langle V\rangle = \mathcal{G}^{-1}R^{-1}V_\mathrm{in}$, which corresponds to the solution of the linear system $Ax = b$ with $A = \mathcal{G}$, $x = V$, $b = R^{-1}V_\mathrm{in}$. 
One may easily add damping through some additional resistors connecting the negative pin and the $V_i$ voltages of each op-amp which has the physical effect of stabilizing the electrical system (as it does numerically), as shown in the scheme with the unnamed resistors.

One may then run the thermodynamic linear solver to solve the system $Ax = b$ by setting the voltage values $V_\mathrm{in}$ to $b/R$ with a digital-to-analog converter, and set the values of the programmable resistors to $R_{ij} = 1/A_{ij}$. In Alg.~\ref{alg:2}, $A = J_k^\top J_k, b = J^\top v_k$ for the $kth$ iteration. 
To obtain the comparisons to other digital methods, we considered the following procedure to run the thermodynamic linear system on electrical hardware:
\begin{enumerate}
    \item Compute $\tilde{v}_k$, and use a Digital-to-analog converter (DAC) to set the $\tilde{v}_k$ vector to voltages $V_{\mathrm{in}}$.
    \item For iteration 0: Set the configuration of the programmable resistors ($(2n+m)^2$ values with a given bit precision to set by calculating $\tilde{J}_0$ digitally. For subsequent iterations, set the diagonals of the blocks to be updated ($4n$ elements with a given bit precision to set).
    \item Let the dynamics run for $t$ (the analog dynamic time). Note that for experiments $t$ was chosen heuristically by exploring convergence in the solutions of the problem of interest, and in general it will be proportional to the relaxation time of the physical system. Since the $\langle V \rangle $ converges exponentially to its steady-state value, this is in general a few time constants (see below for a discussion of this point).
    \item Analog-to-digital (ADC) conversion of the solution measured at nodes $V_i$ to the digital device.
\end{enumerate}
The relaxation time of the system is:
\begin{equation}
    \tau = \frac{RC}{\alpha_{\min}}
\end{equation}

where $R$ is a resistance scale (which means that all resistances  $R_{ij}$  are a multiple of this), $C$ is the capacitance (assuming all the capacitances are the same), and  $\alpha_{\min}$  is the smallest eigenvalue of the (unitless) $\mathcal{G}$  matrix. After this time, all the modes of the system will have relaxed, which may be too conservative (for example, in the case where there is only one slow mode, and all other modes are fast). With regularization,  $\alpha_{\min}$ is lower-bounded by the regularization factor  $\lambda$. For timing purposes, we kept $RC$ as the relaxation time, because of the problem-dependence of  $\alpha_{\min}$.
The runtime estimated are based on the following assumptions:
\begin{itemize}
    \item 16 bits of precision.
    \item A digital transfer speed of $ 100$ Mb/s.
    \item $R = 10^3 \,\Omega$, $C= 1 \,\text{nF}$, which means $RC = 1 \mu \text{s}$ is the characteristic timescale of the system.
\end{itemize}

\bibliographystyle{IEEEtran}

%

\end{document}